\documentclass[12pt]{article}
\usepackage{amsmath}
\usepackage{amssymb}
\begin{document}
\vskip 1truein
\begin{center}
{\bf \Large{Non-consistency of a pure Yang-Mills
\\ type formulation for gravity revisited}}
 \vskip 5pt
{Rolando Gaitan D.}${}^{a
}${\footnote {e-mail: rgaitan@uc.edu.ve}} \\
${}^a${\it Grupo de F\'\i sica Te\'orica, Departamento de F\'\i
sica, Facultad de Ciencias y Tecnolog\'\i a, Universidad de
Carabobo, A.P. 129
Valencia 2001, Edo. Carabobo, Venezuela.}\\
\end{center}
\vskip .2truein
\begin{abstract}


A perturbative regime based on contorsion as a dynamical variable
and metric as a (classical) fixed background, is performed in the
context of a pure Yang-Mills formulation based on $GL(3,R)$ gauge
group. In the massless case we show that the theory propagates
three degrees of freedom and only one is a non-unitary mode. Next,
we introduce quadratical terms dependent on torsion, which
preserve parity and general covariance. The linearized version
reproduces an analogue Hilbert-Einstein-Fierz-Pauli unitary
massive theory plus three massless modes, two of them non-unitary
ones.

\end{abstract}

\vskip .2truein
\section{Introduction}

There were some contributions on the exploration of classical
consistency of a pure Yang-Mills (YM) type formulation for
gravity, including the cosmological extension \cite{g1,g2} (and
the references therein), among others. In those references,
Einstein theory is recovered after the imposition of torsion
constraint.

Unfortunately, the path to a quantum version (if it is finally
possible) is not straightforward. For example, it is well known
that the Lagrangian of a pure YM theory based on the Lorentz group
$SO(3,1)\simeq SL(2,C)$ \cite{kp} leads to a non-positive
Hamiltonian
and, then the canonical quantization procedure fails.
However, there is a possible way out if it is considered an
extension of the YM model thinking about a theory like
Gauss-Bonnet with Torsion (GBT)\cite{kp} and, moreover there
exists a possible family of GBT type theories from which can be
recovered unitarity\cite{g3}.

The aim of this work is to expose, with some detail, a similar
(and obvious) situation about non-unitarity in a YM formulation
with $GL(3,R)$ as a gauge group in both massless and massive
theories. Particularly, the massive version considered here is a
quadratical terms set ($T^2$-terms) preserving parity which
depends on torsion (the old idea about considering $T^2$-terms in
a dynamical theory of torsion has been considered in the
past\cite{hs}) and, at a perturbative regime they give rise to a
Fierz-Pauli massive term. Throughout this work we follow the
spirit of Kibble's idea \cite{kibble} treating the metric as a
fixed background, meanwhile the torsion (contorsion) shall be
considered as a dynamical field and it would be thought as a
quantum fluctuation around a classical fixed background.

This paper is organized as follows. The next section is devoted to
a brief review on notation of the cosmologically extended YM
formulation\cite{g1} in $N$-dimensions and its topologically
massive version in $2+1$\cite{g2}. In section 3, we consider the
scheme of linearization of the massless theory around a fixed
Minkowskian background, allowing fluctuations on torsion. Next,
the Lagrangian analysis of constraints and construction of the
reduced action is performed, showing that this theory does
propagate degrees of freedom, including a ghost. In section 4, we
introduce an appropiate $T^2$-terms, which preserve parity,
general covariance, and its linearization gives rise to a
Fierz-Pauli mass term. There, the non-positive definite
Hamiltonian problem gets worse: the Lagrangian analysis shows that
the theory has more non-unitary degrees of freedom and we can't
expect other thing. Gauge transformations are explored in section
5. Although $T^2$-terms provide mass only to some spin component
of contorsion, the linearized theory loses the gauge invariance
and there is no residual invariance. This is clearly established
through a standard procedure for the study of possible chains of
gauge generators\cite{c}. We end up with some concluding remarks.

\vskip .2truein
\section{A pure Yang-Mills formulation for gravity}

Let $M$ be an $N$-dimensional manifold with a metric, $g_{\mu \nu
}$ provided. A (principal) fiber bundle is constructed with $M$
and a 1-form connection is given,  ${(A_\lambda)^\mu}_\nu $ which
will be non metric dependent. The affine connection transforms as
${A_\lambda}^\prime =UA_\lambda U^{-1} + U
\partial _\lambda U^{-1}$ under $U \in GL(N,R)$. Torsion and curvature tensors are
${T^\mu}_{\lambda\nu}={(A_\lambda)^\mu}_\nu-{(A_\nu)^\mu}_\lambda$
and $F_{\mu\nu} \equiv \partial_\mu A_\nu - \partial_\nu A_\mu +
[A_\mu , A_\nu ]$, respectively. Components of the Riemann tensor
are ${R^\sigma}_{\alpha \mu \nu }\equiv ({F_{\nu \mu})^\sigma
}_\alpha$. The gauge invariant action with cosmological
contribution is\cite{g1}
\begin{equation}
{S^{(N)}}_0=\kappa^{2(4-N)} \big<-\frac 14 \, tr\, F^{\alpha
\beta}F_{\alpha \beta}+ q(N) \lambda^2 \big> \,\, , \label{eqa1}
\end{equation}
where $\kappa^2$ is in length units, $\big<...\big>\equiv \int d^N
x \sqrt{-g}(...) $, $\lambda$ is the cosmologic constant and the
parameter  $q(N)=2(4-N)/(N-2)^2(N-1)$ depends on dimension. The
shape of  $q(N)$ allows the recovering of (free) Einstein's
equations as a particular solution when the torsionless Lagrangian
constraints are imposed and $q(N)$ changes it sign when $N>5$. The
field equations are $\frac 1{\sqrt{-g}}\,\,\partial _\alpha
(\sqrt{-g}\,\,F^{\alpha \lambda}) + [A_\alpha , F^{\alpha
\lambda}] =0$ and ${T_g}^{\alpha\beta}=
-\kappa^2g^{\alpha\beta}\lambda^2$ where
${T_g}^{\alpha\beta}\equiv
\kappa^2\,tr[F^{\alpha\sigma}{F^\beta}_\sigma
-\frac{g^{\alpha\beta}}4 \, F^{\mu \nu}F_{\mu \nu}]$ is the
energy-momentum tensor of gravity.

Particularly, in $2+1$ dimension take place the topologically
massive version\cite{g2}, this means
\begin{equation}
{S^{(3)}}_{tm}={S^{(3)L}}_0 +
\frac{m\kappa^2}{2}\big<\varepsilon^{\mu\nu\lambda}\,tr\big( A_\mu
\partial_\nu A_\lambda +\frac{2}{3}\, A_\mu A_\nu A_\lambda \big)\big>\,
\, , \label{eqa2}
\end{equation}
which contains the cosmologically extended TMG\cite{d} when the
torsionless constraints are introduced through a suitable set of
Lagrangian multipliers. Obviously,  (\ref{eqa2}) does not preserve
parity.

\vskip .2truein
\section{Linearization of the massless theory}
With a view on the performing of a perturbative study of the
massive model, we wish to note some aspects of the variational
analysis of free action (\ref{eqa1}) in $2+1$ dimensions. As we
had said above, the connection shall be considered as  a dynamical
field whereas the space-time metric would be a fixed background,
in order to explore (in some sense) the isolated behavior of
torsion (contorsion) and avoid higher order terms in the field
equations. For simplicity we shall assume $\lambda =0$.

Then, let us  consider a Minkowskian space-time with a metric
$diag(-1,1,1)$ provided and, obviously with no curvature nor
torsion. The notation is
\begin{equation}
\overline{g}_{\alpha\beta}=\eta_{\alpha\beta}\, \, , \label{eqb1}
\end{equation}
\begin{equation}
{\overline{F}}^{\alpha\beta}=0\, \, , \label{eqb2}
\end{equation}
\begin{equation}
{\overline{T}^\lambda}_{\mu\nu}=0\, \, . \label{eqb3}
\end{equation}
It can be observed that curvature ${\overline{F}}^{\alpha\beta}=0$
and torsion ${\overline{T}^\lambda}_{\mu\nu}=0$, in a space-time
with metric $\overline{g}_{\alpha\beta}=\eta_{\alpha\beta}$
satisfy the background equations, $\frac
1{\sqrt{-\overline{g}}}\,\,\partial _\alpha
(\sqrt{-\overline{g}}\,\,\overline{F}^{\alpha \lambda}) +
[\overline{A}_\alpha , \overline{F}^{\alpha \lambda}] =0$ and
${\overline{T}_g}^{\alpha\beta}=0$, identically.

Thinking in variations
\begin{equation}
A_\mu=\overline{A}_\mu+ a_\mu\,\,\,\,,\,\,\,\,|a_\mu|\ll 1\, \, ,
\label{eqb4}
\end{equation}
for this case $\overline{A}_\mu=0$. Then,  action (\ref{eqa1})
takes the form
\begin{equation}
{S^{(3)L}}_0 =\kappa^2 \big<-\frac 14 \, tr\, f^{\alpha
\beta}(a)f_{\alpha \beta}(a)\big> \,\, , \label{eqb5}
\end{equation}
where $f_{\alpha \beta}(a)=\partial_\alpha a_\beta-\partial_\beta
a_\alpha $ and (\ref{eqb5}) is gauge invariant under
\begin{equation}
\delta a_\mu = \partial_\mu \omega  \,\, , \label{eqb6}
\end{equation}
with the gauge group
$G=U(1)\times...{3^2}...\times U(1)$.

Let us suppose we take a $Weitzenb\ddot{o}ck$ space-time instead a
Minkowski one, as the fixed background. Then, the condition
(\ref{eqb3}) must be relaxed (i.e.:
${\overline{T}^\lambda}_{\mu\nu}\neq 0$) and the linearized action
would be $ S^{(3)L}_{Weitzenb\ddot{o}ck}= {S^{(3)L}}_0 -\kappa^2
\big<tr f^{\alpha \beta}(a)[\overline{A}_\alpha , a_\beta] +\frac
12 \, tr\, [\overline{A}^\alpha ,
a^\beta]\big([\overline{A}_\alpha , a_\beta]-[\overline{A}_\beta ,
a_\alpha]\big) \big>$, which now is gauge variant under
(\ref{eqb6}). In this context, the gauge invariance can be
recovered through a similar technique of $St\ddot{u}ckelberg$
auxilliary fields for a $Proca$ type model. This situation
suggests we would be in presence of possible massive contributions
(surely with ghosts) due just to the background.

In order to describe in detail the action (\ref{eqb5}), let us
consider the following decomposition for perturbed connection
\begin{equation}
{(a_\mu)^{\alpha}}_\beta={\epsilon^{\sigma\alpha}}_\beta
k_{\mu\sigma}+{\delta^\alpha}_\mu v_\beta
-\eta_{\mu\beta}v^\alpha\,\, , \label{eqb7}
\end{equation}
where $k_{\mu\nu}=k_{\nu\mu}$ and $v_\mu$ are the symmetric and
antisymmetric parts of the rank two perturbed contorsion  (i. e.,
the rank two contorsion is $K_{\mu\nu}\equiv -\frac{1}2
{\epsilon^{\sigma\rho}}_\nu K_{\sigma\mu\rho}$), respectively. It
can be noted that decomposition  (\ref{eqb7}) has not been
performed in irreducible spin components and explicit writing down
of the traceless part of $k_{\mu\nu}$ would be needed. This
component will be considered when the study of reduced action
shall be performed. Using (\ref{eqb7}) in (\ref{eqb5}), we get
\begin{equation}
{S^{(3)L}}_0 =\kappa^2 \big< k_{\mu\nu}\Box{}k^{\mu\nu}+
\partial_\mu k^{\mu\sigma}\partial_\nu {k^\nu}_\sigma
-2 \epsilon^{\sigma\alpha\beta}\partial_\alpha v_\beta\partial_\nu
{k^\nu}_\sigma - v_\mu \Box{}v^\mu +(\partial_\mu v^\mu)^2\big>
\,\, , \label{eqb8}
\end{equation}
which is gauge invariant under the following transformation rules
(induced by (\ref{eqb6}))
\begin{equation}
\delta k_{\mu\nu}= \partial_\mu \xi_\nu + \partial_\nu \xi_\mu
\,\, , \label{eqb9}
\end{equation}
\begin{equation}
\delta v_\mu =-{\epsilon^{\sigma\rho}}_\mu \partial_\sigma
\xi_\rho \,\, , \label{eqb10}
\end{equation}
with $\xi_\mu \equiv \frac{1}4
{\epsilon^\beta}_{\alpha\mu}{w^\alpha}_\beta$. These
transformation rules clearly show that only the antisymmetric part
of $w$ is needed (i. e.: only three gauge fixation would be
chosen).

In expression (\ref{eqb8}) we can observe that the term $v_\mu
\Box{}v^\mu$ has a wrong sign, telling us about the non-unitarity
property of the theory. However, field equations are
\begin{equation}
2\Box{}k_{\mu\nu}-
\partial_\mu\partial_\sigma {k^\sigma}_\nu-\partial_\nu\partial_\sigma {k^\sigma}_\mu
+{\epsilon^{\sigma\rho}}_\mu\partial_\nu\partial_\sigma v_\rho
+{\epsilon^{\sigma\rho}}_\nu\partial_\mu\partial_\sigma v_\rho =0
\,\, , \label{eqb11}
\end{equation}
\begin{equation}
\epsilon^{\sigma\rho\beta}\partial_\sigma \partial_\mu
{k^\mu}_\rho +\Box{}v^\beta +\partial^\beta\partial_\mu v^\mu=0
\,\, , \label{eqb12}
\end{equation}
and note that  (\ref{eqb11}) satisfies the consistency condition
\begin{equation}
\Box{}k-\partial_\mu\partial_\nu k^{\mu\nu}=0 \,\,. \label{eqb12a}
\end{equation}

Divergence of (\ref{eqb12}) says that $\partial_\mu v^\mu$ is a
massless 0-form then, if we define $\hat{\partial}_\sigma\equiv
\Box{}^{-\frac{1}2}\partial_\sigma $, the following relation can
be written
\begin{equation}
v^\beta=-\epsilon^{\sigma\rho\beta}\hat{\partial}_\sigma
\hat{\partial}_\mu {k^\mu}_\rho  \,\, , \label{eqb13}
\end{equation}
up to a massless-transverse 1-form. Using  (\ref{eqb13}) in
(\ref{eqb11}), gives rise to
\begin{equation}
\Box{}k_{\mu\nu}-
\partial_\mu\partial_\sigma {k^\sigma}_\nu-\partial_\nu\partial_\sigma {k^\sigma}_\mu
+\partial_\mu\partial_\nu k=0 \,\, , \label{eqb14}
\end{equation}
up to a massless 0-form. This last equation with condition
(\ref{eqb12a}) would suggests a possible equivalence with the
model for gravitons of the linearized Hilbert-Einstein theory (i.
e.: free gravity in $2+1$ does not propagate degrees of freedom).
However, this suggestion is wrong because we were dropped out too
many light modes and, then it is necessary to take into account
both massive and massless complete sets of modes (light modes are
relevant at the lower energy regime).

Now, let us study the system of Lagrangian constraints in order to
explore the number of degrees of freedom. A possible approach
consists in a $2+1$ decomposition of the action (\ref{eqb8}) in
the way
\begin{eqnarray}
{S^{(3)L}}_{0} =\kappa^2
\big<[-\dot{k}_{0i}+2\partial_ik_{00}-2\partial_nk_{ni}-2\epsilon_{in}\dot{v}_n
+2\epsilon_{in}\partial_n v_0
]\dot{k}_{0i}+\dot{k}_{ij}\dot{k}_{ij}\nonumber \\
+[2\epsilon_{nj}\partial_nk_{00}+2\epsilon_{nj}\partial_mk_{nm}-\dot{v}_j-2\partial_jv_0]\dot{v}_j+2(\dot{v}_0)^2
+k_{00}\Delta k_{00}\nonumber \\-2k_{0i}\Delta k_{0i}+k_{ij}\Delta
k_{ij}-(\partial_ik_{i0})^2+\partial_nk_{ni}\partial_mk_{mi}
-2\epsilon_{ij}\partial_i v_j
\partial_nk_{n0}\nonumber \\-2\epsilon_{lm}\partial_m v_0
\partial_nk_{nl}+v_0\Delta v_0-v_i\Delta v_i +(\partial_nv_n)^2\big>\,\,,
\label{eqb14a}
\end{eqnarray}
and using a Transverse-Longitudinal (TL) decomposition\cite{pio}
with notation
\begin{eqnarray}
k_{00}\equiv n \, \, , \label{a1}
\end{eqnarray}
\begin{eqnarray}
h_{i0}=h_{0i} \equiv \partial_i k^L + \epsilon_{il}\partial_l
k^T\, \, , \label{a2}
\end{eqnarray}
\begin{eqnarray}
k_{ij}=k_{ji}\equiv (\eta_{ij}\Delta -\partial _i \partial
_j)k^{TT} +\partial _i \partial _j k^{LL} +(\epsilon _{ik}\partial
_k\partial _j +\epsilon _{jk}\partial _k \partial _i )k^{TL}\, \,
, \label{a3}
\end{eqnarray}
\begin{eqnarray}
v_0\equiv q \, \, , \label{a4}
\end{eqnarray}
\begin{eqnarray}
v_i \equiv \partial_i v^L + \epsilon_{il}\partial_l v^T\, \, ,
\label{a5}
\end{eqnarray}
where $\Delta \equiv \partial _i \partial _i$, eq. (\ref{eqb14a})
 can be rewritten as follows
\begin{eqnarray}
{S^{(3)L}}_{0} =\kappa^2 \big<\dot{k}^L \Delta
\dot{k}^L+\dot{k}^T \Delta \dot{k}^T+\dot{v}^L \Delta \dot{v}^L+
\dot{v}^T \Delta \dot{v}^T+2\dot{v}^L \Delta \dot{k}^T-2\dot{v}^T
\Delta \dot{k}^L
\nonumber \\
+(\Delta \dot{k}^{TT})^2 +(\Delta \dot{k}^{LL})^2 +2(\Delta
\dot{k}^{TL})^2+2(\dot{q})^2-2n\Delta \dot{k}^L +2n\Delta
\dot{v}^T\nonumber
\\+2q\Delta \dot{v}^L-2q\Delta
\dot{k}^T+2\Delta k^{LL}\Delta \dot{k}^L+2\Delta k^{TL}\Delta
\dot{k}^T+2\Delta k^{LL}\Delta \dot{v}^T \nonumber \\-2\Delta
k^{TL}\Delta \dot{v}^L +q\Delta q+n\Delta n+(\Delta
k^L)^2+2(\Delta k^T)^2 +2(\Delta v^L)^2\nonumber \\ +(\Delta
v^T)^2+2\Delta v^{T}\Delta k^{L} +2q\Delta^2 k^{TL} +\Delta
k^{TT}\Delta^2 k^{TT}+\Delta k^{TL}\Delta^2 k^{TL}\big>\,\,.
\label{e14}
\end{eqnarray}

Primary Lagrangian constraints, joined to some links among
accelerations, can be obtained through an inspection on field
equations, which arise from (\ref{e14}). A ''Coulomb'' gauge is
defined by the constraints $\partial_ik_{i\mu}=0$, which can be
rewritten in terms of the TL-decomposition as follows (up to
harmonics)
\begin{equation}
k^L=k^{LL}=k^{TL}=0 \,\, , \label{e15}
\end{equation}
and preservation provides the next conditions for longitudinal
velocities and accelerations
\begin{equation}
\dot{k}^L=\dot{k}^{LL}=\dot{k}^{TL}=0 \,\, , \label{e16}
\end{equation}
\begin{equation}
\ddot{k}^L=\ddot{k}^{LL}=\ddot{k}^{TL}=0 \,\, . \label{e17}
\end{equation}
Equations (\ref{e15}) and (\ref{e16}) are six Lagrangian
constraints.

Field equations with the help of gauge constraints, give the
following  five (primary) constraints
\begin{equation}
n=0 \,\, , \label{e18}
\end{equation}
\begin{equation}
v^T=0 \,\, , \label{e19}
\end{equation}
\begin{equation}
\dot{v}^T=0 \,\, , \label{e20}
\end{equation}
\begin{equation}
\dot{k}^T-\dot{v}^L+q=0 \,\, , \label{e21}
\end{equation}
\begin{equation}
\Delta k^T-\Delta v^L+\dot{q}=0 \,\, , \label{e22}
\end{equation}
and accelerations are related through
\begin{equation}
\ddot{v}^T=-\dot{n} \,\, , \label{e23}
\end{equation}
\begin{equation}
\ddot{k}^T+\ddot{v}^L=\dot{q} +2\Delta {k}^T\,\, , \label{e24}
\end{equation}
\begin{equation}
\ddot{k}^{TT}=\Delta k^{TT} \,\, , \label{e25}
\end{equation}
\begin{equation}
\ddot{q}=\Delta q \,\, . \label{e26}
\end{equation}

Systematic preservation of constraints (\ref{e18}) and (\ref{e22}) provide a new constraint
\begin{equation}
\dot{n}=0 \,\, , \label{e27}
\end{equation}
and accelerations
\begin{equation}
\ddot{n}=0 \,\, , \label{e28}
\end{equation}
\begin{equation}
\ddot{v}^T=0 \,\, , \label{e29}
\end{equation}
\begin{equation}
\ddot{k}^T=\Delta {k}^T\,\, , \label{e30}
\end{equation}
\begin{equation}
\ddot{v}^L=\Delta {v}^L \,\, . \label{e31}
\end{equation}

In short, there is a set of twelve constraints
\begin{equation}
n=\dot{n}=v^T=\dot{v}^T=k^L=\dot{k}^L=k^{LL}=\dot{k}^{LL}=k^{TL}=\dot{k}^{TL}=0 \,\, , \label{e32}
\end{equation}
\begin{equation}
\dot{k}^T-\dot{v}^L+q=0 \,\, , \label{e33}
\end{equation}
\begin{equation}
\Delta k^T-\Delta v^L+\dot{q}=0 \,\, , \label{e34}
\end{equation}
then, there are three degrees of freedom, and the constraint
system give rise to reduced action
\begin{eqnarray}
{S^{(3)L*}}_{0} =\kappa^2 \big<4\dot{k}^T \Delta \dot{k}^T+4(\Delta k^T)^2
+4(\dot{q})^2+4q\Delta q
+(\Delta \dot{k}^{TT})^2+\Delta
k^{TT}\Delta^2 k^{TT}\big>\,\,.
\label{e35}
\end{eqnarray}
Introducing notation
\begin{equation}
Q\equiv 2q \,\, , \label{e36}
\end{equation}
\begin{equation}
Q^T\equiv 2(-\Delta)^{\frac{1}2}k^T \,\, , \label{e37}
\end{equation}
\begin{equation}
Q^{TT}\equiv \Delta k^{TT} \,\, , \label{e38}
\end{equation}
the reduced action is rewritten as follows
\begin{eqnarray}
{S^{(3)L*}}_{0} =\kappa^2 \big<Q\Box{}Q-Q^T\Box{}Q^T+ Q^{TT}\Box{}Q^{TT}\big>\,\,,
\label{e35}
\end{eqnarray}
showing two unitary and one non-unitary modes, then the
Hamiltonian is not positive definite. This study could also have
considered from the point of view of the exchange amplitude
procedure, in which is considered the coupling to a (conserved)
energy-momentum tensor of some source, trough Lagrangian terms
$\kappa k_{\mu\nu}T^{\mu\nu}$ and $\chi v_{\mu}J^{\mu}$.

\vskip .2truein
\section{YM gravity with parity preserving massive term}

It can be possible to write down a massive version
which respect parity, for example
\begin{equation}
{S^{(3)}}_{m}={S^{(3)}}_0 - \frac{m^2
\kappa^2}{2}\big<{T^{\sigma}}_{\sigma
\nu}{{T^{\rho}}_{\rho}}^{\nu}-T^{\lambda\mu\nu}T_{\mu\lambda\nu}-\frac{1}{2}T^{\lambda\mu\nu}T_{\lambda\mu\nu}\big>\,
\, . \label{eqa3}
\end{equation}
In a general case, if we allow independent variations on metric and connection  two
types of field equations can be obtained. On one hand, variations
on metric give rise to the expression of the gravitacional
energy-momentum tensor, ${T_g}^{\alpha\beta}\equiv
\kappa^2\,tr[F^{\alpha\sigma}{F^\beta}_\sigma
-\frac{g^{\alpha\beta}}4 \, F^{\mu \nu}F_{\mu \nu}]$, in other
words
\begin{equation}
{T_g}^{\alpha\beta}=
-{T_t}^{\alpha\beta}-\kappa^2g^{\alpha\beta}\lambda^2\, \, ,
\label{eqa4}
\end{equation}
where ${T_t}^{\alpha\beta}\equiv
-m^2\kappa^2[3t^{\alpha\sigma}{t^\beta}_\sigma
+3t^{\sigma\alpha}{t_\sigma}^\beta
-t^{\alpha\sigma}{t_\sigma}^\beta-t^{\sigma\alpha}{t^\beta}_\sigma
- (t^{\alpha\beta}+t^{\beta\alpha}){t_\sigma}^\sigma
-\frac{5g^{\alpha\beta}}2 \, t^{\mu \nu}t_{\mu
\nu}+\frac{3g^{\alpha\beta}}2 \, t^{\mu \nu}t_{\nu
\mu}+\frac{g^{\alpha\beta}}2\,({t_\sigma}^\sigma)^2 ]$ is the
torsion contribution to the energy-momentum distribution and
$t^{\alpha\beta}\equiv\frac{\varepsilon^{\mu\nu\alpha}}2\,{T^{\beta}}_{\mu\nu}
$. This says, for example, that the quest of possible black hole
solutions must reveal a dependence on parameters  $m^2$ and
$\lambda^2$.

On the other hand, variations on connection provide the following
equations
\begin{eqnarray}
\frac 1{\sqrt{-g}}\,\,\partial _\alpha (\sqrt{-g}\,\,F^{\alpha
\lambda}) + [A_\alpha , F^{\alpha \lambda}] =J^\lambda \, \, ,
\label{eqa5}
\end{eqnarray}
where the current is $(J^\lambda)^{\nu}\,_\sigma =
m^2({\delta^\lambda}_\sigma {{K^\rho}_\rho}^\nu
-{\delta^\nu}_\sigma {{K^\rho}_\rho}^\lambda
+2{{K^\nu}_\sigma}^\lambda)$ and the contorsion
${K^\lambda}_{\mu\nu}\equiv \frac
1{2}({T^\lambda}_{\mu\nu}+{{T_\mu}^\lambda}_\nu +
{{T_\nu}^\lambda}_\mu)$. We can observe in (\ref{eqa5}) that
contorsion and metric appear as sources of gravity, where the
cosmological contribution is obviously hide in space-time metric.
In a weak torsion regime, equation (\ref{eqa5}) takes a familiar
shape: $D_\alpha F^{\alpha \lambda} =J^\lambda $, where $D_\alpha
$ is the covariant derivative computed with the Christoffel's
symbols.

Now we explore the perturbation of the massive case given at
(\ref{eqa3}) and with the help of (\ref{eqb7}), the linearized
action is
\begin{eqnarray}
{S^{(3)L}}_m =\kappa^2 \big< k_{\mu\nu}\Box{}k^{\mu\nu}+
\partial_\mu k^{\mu\sigma}\partial_\nu {k^\nu}_\sigma
-2 \epsilon^{\sigma\alpha\beta}\partial_\alpha v_\beta
\partial_\nu {k^\nu}_\sigma\nonumber \\ - v_\mu \Box{}v^\mu +(\partial_\mu
v^\mu)^2-m^2(k_{\mu\nu}k^{\mu\nu}-k^2)\big> \,\, . \label{eqb15}
\end{eqnarray}

Using a TL-decomposition defined by (\ref{a1})-(\ref{a5}), we can
write  (\ref{eqb15}) in the way
\begin{eqnarray}
{S^{(3)L}}_m =\kappa^2 \big<\dot{k}^L \Delta \dot{k}^L+\dot{k}^T
\Delta \dot{k}^T+\dot{v}^L \Delta \dot{v}^L+ \dot{v}^T \Delta
\dot{v}^T+2\dot{v}^L \Delta \dot{k}^T-2\dot{v}^T \Delta \dot{k}^L
\nonumber \\
+(\Delta \dot{k}^{TT})^2 +(\Delta \dot{k}^{LL})^2 +2(\Delta
\dot{k}^{TL})^2+2(\dot{q})^2-2n\Delta \dot{k}^L +2n\Delta
\dot{v}^T\nonumber
\\+2q\Delta \dot{v}^L-2q\Delta
\dot{k}^T+2\Delta k^{LL}\Delta \dot{k}^L+2\Delta k^{TL}\Delta
\dot{k}^T+2\Delta k^{LL}\Delta \dot{v}^T \nonumber \\-2\Delta
k^{TL}\Delta \dot{v}^L +q\Delta q+n\Delta n+(\Delta
k^L)^2+2(\Delta k^T)^2 +2(\Delta v^L)^2\nonumber \\ +(\Delta
v^T)^2+2\Delta v^{T}\Delta k^{L} +2q\Delta^2 k^{TL} +\Delta
k^{TT}\Delta^2 k^{TT}+\Delta k^{TL}\Delta^2 k^{TL}\nonumber
\\+m^2[-2k^L\Delta k^L-2k^T\Delta k^T-2(\Delta  k^{TL})^2 -2n(\Delta  k^{TT}+\Delta  k^{LL})\nonumber
\\+2\Delta  k^{TT}\Delta  k^{LL}]\big>\,\,. \label{eqba1}
\end{eqnarray}
Here, there is no gauge freedom (as it shall be confirmed in next
section) and field equations provide primary constraints and some
accelerations. The preservation procedure gives rise to
expressions for all accelerations
\begin{equation}
\ddot{n}=\Delta \dot{k}^L \,\, , \label{eqba2}
\end{equation}
\begin{equation}
\ddot{k}^L=-\Delta \dot{k}^{TT}+\dot{n} \,\, , \label{eqba3}
\end{equation}
\begin{equation}
\ddot{k}^T=\Delta \dot{k}^{TL}\,\, , \label{eqba4}
\end{equation}
\begin{equation}
\ddot{k}^{LL}= \dot{k}^L+ \dot{v}^T+m^2k^{TT}-m^2\Delta^{-1} n
\,\, , \label{eqba5}
\end{equation}
\begin{equation}
\ddot{k}^{TL}=\frac{1}2 (\dot{k}^T- \dot{v}^L+\Delta
k^{TL}+q-2m^2k^{TL}) \,\, , \label{eqba6}
\end{equation}
\begin{equation}
\ddot{k}^{TT}=\Delta k^{TT}+m^2k^{LL}-m^2\Delta^{-1} n\,\, ,
\label{eqba7}
\end{equation}
\begin{equation}
\ddot{q}=\frac{1}2 (\Delta\dot{v}^L-\Delta \dot{k}^T+\Delta^2
k^{TL}+\Delta q) \,\, , \label{eqba8}
\end{equation}
\begin{equation}
\ddot{v}^L=-\dot{q}+2\Delta v^L\,\, , \label{eqba9}
\end{equation}
\begin{equation}
\ddot{v}^T=-\Delta \dot{k}^{TT}-\Delta \dot{k}^{LL}+\Delta
k^L+\Delta v^T \,\, , \label{eqba10}
\end{equation}
and a set of eight constraints
\begin{equation}
\dot{v}^T-\dot{k}^L+n -m^2(k^{TT}+k^{LL})=0\,\, , \label{c1}
\end{equation}
\begin{equation}
\Delta \dot{k}^{LL}-\Delta k^L-\Delta v^T +m^2k^L=0\,\, ,
\label{c2}
\end{equation}
\begin{equation}
\Delta \dot{k}^{TL}-\dot{q}-\Delta k^T+\Delta v^L +m^2k^T=0 \,\, ,
\label{c3}
\end{equation}
\begin{equation}
\Delta \dot{k}^{LL}-\Delta k^L-\Delta v^T
+m^2(\dot{k}^{TT}+\dot{k}^{LL})=0\,\, , \label{c4}
\end{equation}
\begin{equation}
\dot{k}^L+\Delta k^{TT}-n =0 \,\, , \label{c5}
\end{equation}
\begin{equation}
\dot{k}^T-\Delta k^{TL}=0\,\, , \label{c6}
\end{equation}
\begin{equation}
\dot{v}^T+\Delta k^{TT}+m^2(k^{TT}+k^{LL})-2m^2\Delta^{-1} n=0
\,\, , \label{c7}
\end{equation}
\begin{equation}
\dot{n}-\Delta k^{L}=0 \,\, , \label{c8}
\end{equation}
which says that this massive theory propagates five degrees of
freedom. In order to explore the physical content, we can take a
short path to this purpose and it means to start with a typical
transverse-traceless (Tt) decomposition instead the
TL-decomposition one. Notation for the Tt-decomposition of fields
is
\begin{equation}
k_{\mu\nu}={k^{Tt}}_{\mu\nu}+\hat{\partial}_\mu {\theta^{T}}_\nu
+\hat{\partial}_\nu
{\theta^{T}}_\mu+\hat{\partial}_\mu\hat{\partial}_\nu \psi
+\eta_{\mu\nu}\phi \,\, , \label{eqb16}
\end{equation}
\begin{equation}
v_\mu={v^{T}}_\mu+\hat{\partial}_\mu v \,\, , \label{eqb17}
\end{equation}
with the subsidiary conditions
\begin{equation}
{k^{Tt\mu}}_\mu=0\,\,\,\,,\,\,\,\,\partial^\mu{k^{Tt}}_{\mu\nu}=0\,\,\,\,,\,\,\,\,\partial^\mu
{\theta^{T}}_\mu =0\,\,\,\,,\,\,\,\,\partial^\mu {v^{T}}_\mu =0
\,\, . \label{eqb18}
\end{equation}
Action (\ref{eqb15}) is
\begin{eqnarray}
{S^{(3)L}}_m =\kappa^2 \big<
{k^{Tt}}_{\mu\nu}(\Box{}-m^2){k^{Tt}}^{\mu\nu}-{\theta^{T}}_\mu(\Box{}-2m^2){\theta^{T}}_\mu
-2 \epsilon^{\sigma\alpha\beta}\partial_\alpha {v^{T}}_\beta
\Box{}^{\frac{1}2}{\theta^{T}}_\sigma\nonumber \\ - {v^{T}}_\mu
\Box{}{v^{T}}^\mu
+2v\Box{}v+2\phi\Box{}\phi+4m^2\psi\phi+6m^2\phi^2\big> \,\, .
\label{d1}
\end{eqnarray}
A new transverse variable, ${a^T}_\mu$ is introduced through
\begin{equation}
{\theta^{T}}_\mu \equiv
{\epsilon_\mu}^{\alpha\beta}\hat{\partial}_\alpha
{a^{T}}_\beta\,\, , \label{d2}
\end{equation}
and the action (\ref{d1}) is rewritten as
\begin{eqnarray}
{S^{(3)L}}_m =\kappa^2 \big<
{k^{Tt}}_{\mu\nu}(\Box{}-m^2){k^{Tt}}^{\mu\nu}-{a^{T}}_\mu(\Box{}-2m^2){a^{T}}_\mu
-2 {a^{T}}_\mu\Box{}{v^{T}}^\mu \nonumber \\ - {v^{T}}_\mu
\Box{}{v^{T}}^\mu
+2v\Box{}v+2\phi\Box{}\phi+4m^2\psi\phi+6m^2\phi^2\big> \,\, .
\label{d3}
\end{eqnarray}
The field equations are
\begin{equation}
(\Box{}-m^2){k^{Tt}}_{\mu\nu}=0 \,\, , \label{d4}
\end{equation}
\begin{equation}
\Box{} {v^{T}}_\mu=0 \,\, , \label{d5}
\end{equation}
\begin{equation}
\Box{}v=0\,\, , \label{d6}
\end{equation}
\begin{equation}
{a^{T}}_\mu=0 \,\, , \label{d7}
\end{equation}
\begin{equation}
 \psi=\phi=0 \,\, , \label{d8}
\end{equation}
and reduced action is
\begin{eqnarray}
{S^{(3)L*}}_m =\kappa^2 \big<
{k^{Tt}}_{\mu\nu}(\Box{}-m^2){k^{Tt}}^{\mu\nu}+2v\Box{}v-
{v^{T}}_\mu \Box{}{v^{T}}^\mu \big> \,\, , \label{d3}
\end{eqnarray}
saying that the contorsion propagates two massive helicities $\pm
2$, one massless spin-$0$ and two massless ghost vectors. Then,
there is not positive definite Hamiltonian. This observation can
be confirmed in the next section when we shall write down the
Hamiltonian density and a wrong sign appears in the kinetic part
corresponding to the canonical momentum of $v_i$ (see eq.
(\ref{eqb33})).

\vskip .2truein
\section{Gauge transformations}

The quadratical Lagrangian density dependent in torsion and
presented in (\ref{eqa3}), has been constructed without free
parameters, with the exception of $m^2$, of course. It has a
particular shape which only gives mass to the spin 2 component of
the contorsion, as we see in the perturbative regime. Let us
comment about de non existence of any possible ''residual'' gauge
invariance of the model. The answer is that the model lost its
gauge invariance and it can be shown performing the study of
symmetries through computation of the gauge generator chains. For
this purpose, a $2+1$ decomposition of (\ref{eqb15}) is performed,
this means
\begin{eqnarray}
{S^{(3)L}}_{m} =\kappa^2
\big<[-\dot{k}_{0i}+2\partial_ik_{00}-2\partial_nk_{ni}-2\epsilon_{in}\dot{v}_n
+2\epsilon_{in}\partial_n v_0
]\dot{k}_{0i}+\dot{k}_{ij}\dot{k}_{ij}\nonumber \\
+[2\epsilon_{nj}\partial_nk_{00}+2\epsilon_{nj}\partial_mk_{nm}-\dot{v}_j-2\partial_jv_0]\dot{v}_j+2(\dot{v}_0)^2
+k_{00}\Delta k_{00}\nonumber \\-2k_{0i}\Delta k_{0i}+k_{ij}\Delta
k_{ij}-(\partial_ik_{i0})^2+\partial_nk_{ni}\partial_mk_{mi}
-2\epsilon_{ij}\partial_i v_j
\partial_nk_{n0}\nonumber \\-2\epsilon_{lm}\partial_m v_0
\partial_nk_{nl}+v_0\Delta v_0-v_i\Delta v_i +(\partial_nv_n)^2\nonumber \\
+m^2[2k_{0i}k_{0i}-k_{ij}k_{ij}-2k_{00}k_{ii}+(k_{ii})^2]\big>\,\,,
\label{eqb22}
\end{eqnarray}
where $\epsilon_{ij}\equiv {\epsilon^0}_{ij}$.

Next, the  momenta are
\begin{equation}
\Pi\equiv \frac{\partial\mathcal{L}}{\partial\dot{k}_{00}}=0\,\, ,
\label{eqb23}
\end{equation}
\begin{equation}
\Pi^i\equiv
\frac{\partial\mathcal{L}}{\partial\dot{k}_{0i}}=-2\dot{k}_{0i}-2\epsilon_{in}
\dot{v}_n+2\partial_ik_{i0}-2\partial_nk_{ni}+2\epsilon_{in}
\partial_nv_0\,\, , \label{eqb24}
\end{equation}
\begin{equation}
\Pi^{ij}\equiv
\frac{\partial\mathcal{L}}{\partial\dot{k}_{ij}}=2\dot{k}_{ij}\,\,
, \label{eqb25}
\end{equation}
\begin{equation}
P\equiv
\frac{\partial\mathcal{L}}{\partial\dot{v}_0}=4\dot{v}_0\,\, ,
\label{eqb26}
\end{equation}
\begin{equation}
P^j\equiv
\frac{\partial\mathcal{L}}{\partial\dot{v}_j}=-2\epsilon_{nj}\dot{k}_{0n}-2\dot{v}_j+2\epsilon_{nj}\partial_nk_{00}
+2\epsilon_{nj}\partial_mk_{mn} -2\partial_jv_0\,\, ,
\label{eqb27}
\end{equation}
and we establish the following commutation rules
\begin{equation}
\{k_{00}(x),\Pi (y)\}=\{v_0(x),P(y)\}=\delta^2(x-y)\,\, ,
\label{eqb28}
\end{equation}
\begin{equation}
\{k_{0i}(x),\Pi^j (y)\}=\{v_i(x),P^j(y)\}={\delta^j}_i
\delta^2(x-y)\,\, , \label{eqb29}
\end{equation}
\begin{equation}
\{k_{ij}(x),\Pi^{nm}(y)\}=\frac{1}2({\delta^n}_i{\delta^m}_j+{\delta^m}_i{\delta^n}_j)
\delta^2(x-y)\,\,. \label{eqb30}
\end{equation}

It can be noted that (\ref{eqb23}) is a primary constraint that we
name
\begin{equation}
G^{(K)}\equiv\Pi\,\, , \label{eqb31}
\end{equation}
where $K$ means the initial index corresponding to a possible
gauge generator chain, provided by the algorithm developed in
reference\cite{c}. Moreover, manipulating (\ref{eqb25}) and
(\ref{eqb27}), other primary constraints appear
\begin{equation}
G^{(K)}_i\equiv \partial_n
k_{ni}-\epsilon_{in}\partial_nv_0-\frac{\epsilon_{in}}{4}P^n+\frac{1}{4}\Pi^i\,\,
, \label{eqb32}
\end{equation}
and we observe that $G^{(K)}$ and $G^{(K)}_i$ are first class.

The preservation of constraints requires to obtain the Hamiltonian
of the model. First of all, the Hamiltonian density can be written
as
$\mathcal{H}_0=\Pi^i\dot{h}_{0i}+\Pi^{ij}\dot{h}_{ij}+P\dot{v}_0+P^i\dot{v}_i-\mathcal{L}$,
in other words
\begin{eqnarray}
\mathcal{H}_0 =\frac{\Pi^{ij}\Pi^{ij}}{4}+ \frac{P^2}{8}
-\frac{P^iP^i}{4}+\epsilon_{nj}\partial_mk_{nm}P^j+v_0[\partial_iP^i+4\epsilon_{ml}\partial_m\partial_nk_{nl}]\nonumber
\\+k_{00}[2\partial_m\partial_nk_{mn}-\epsilon_{nm}\partial_nP^m+2m^2k_{ii}]
+2k_{0i}\Delta k_{0i}-k_{ij}\Delta k_{ij}\nonumber
\\+(\partial_ik_{i0})^2-2\partial_nk_{ni}\partial_mk_{mi}
+2\epsilon_{ij}\partial_i v_j
\partial_nk_{n0}+v_i\Delta v_i -(\partial_nv_n)^2\nonumber \\
-m^2[2k_{0i}k_{0i}-k_{ij}k_{ij}+(k_{ii})^2]\,\,. \label{eqb33}
\end{eqnarray}

Then, the Hamiltonian is $H_0=\int dy^2 \mathcal{H}_0(y)\equiv
\big<\mathcal{H}_0\big>_y$ and the preservation of $G^{(K)}$,
defined in (\ref{eqb31}) is
\begin{equation}
\{G^{(K)}(x),H_0\}=-2\partial_m\partial_nk_{mn}(x)+\epsilon_{nm}\partial_nP^m(x)-2m^2k_{ii}(x)\,\,
. \label{eqb34}
\end{equation}
The possible generators chain is given by the rule:
''$G^{(K-1)}+\{G^{(K)}(x),H_0\}=${\it combination of primary
constraints}'', then
\begin{eqnarray}
G^{(K-1)}(x)=2\partial_m\partial_nk_{mn}(x)-\epsilon_{nm}\partial_nP^m(x)+2m^2k_{ii}(x)
\nonumber \\+\big<a(x,y)G^{(K)}(y)+b^i(x,y)G^{(K)}_i(y)\big>_y\,\,
. \label{eqb35}
\end{eqnarray}

The preservation of $G^{(K)}_i$, defined in (\ref{eqb32}), is
\begin{eqnarray}
\{G^{(K)}_i(x),H_0\}=\frac{\partial_n\Pi^{ni}(x)}{2}-\frac{\epsilon_{in}}{4}\partial_nP(x)+\frac{\epsilon_{in}}{2}\Delta
v_n(x) +\frac{\epsilon_{in}}{2}\partial_n\partial_mv_m(x)\nonumber
\\+\frac{\epsilon_{nm}}{2}\partial_i\partial_nv_m(x)-(\Delta
-m^2)k_{0i}(x) \,\, , \label{eqb36}
\end{eqnarray}
then
\begin{eqnarray}
G^{(K-1)}_i(x)=-\frac{\partial_n\Pi^{ni}(x)}{2}+\frac{\epsilon_{in}}{4}\partial_nP(x)-\frac{\epsilon_{in}}{2}\Delta
v_n(x) -\frac{\epsilon_{in}}{2}\partial_n\partial_mv_m(x)\nonumber
\\-\frac{\epsilon_{nm}}{2}\partial_i\partial_nv_m(x)+(\Delta
-m^2)k_{0i}(x)\nonumber
\\
+\big<a^i(x,y)G^{(K)}(y)+{b^i}_j(x,y)G^{(K)}_j(y)\big>_y\,\, .
\label{eqb37}
\end{eqnarray}

The undefined objects $a(x,y)$, $b^i(x,y)$, $a^i(x,y)$ and
${b^i}_j(x,y)$ in  expressions (\ref{eqb35}) and (\ref{eqb37}),
are functions or distributions. If it is possible, they can be
fixed in a way that the preservation of  $G^{(K-1)}(x)$ and
$G^{(K-1)}_i(x)$ would be combinations of primary constraints.
With this, the generator chains could be interrupted and we simply
take $K=1$. Of course, the order $K-1=0$ generators must be first
class, as every one. Next, we can see that all these statements
depend on the massive or non-massive character of the theory.

Taking a chain with $K=1$, the candidates to generators of gauge
transformation are (\ref{eqb31}), (\ref{eqb32}), (\ref{eqb35}) and
(\ref{eqb37}). But, the only non null commutators are
\begin{eqnarray}
\{G^{(1)}_i(x),G^{(0)}_j(y)\}=\frac{m^2}{4}\eta_{ij}\delta^2(x-y)
\,\, , \label{eqb38}
\end{eqnarray}
\begin{eqnarray}
\{G^{(0)}(x),G^{(0)}_i(y)\}=m^2\big(\partial_i\delta^2(x-y)+\frac{b^i(x,y)}{4}\big)
\,\, , \label{eqb39}
\end{eqnarray}
saying that the system of ''generators'' is not first class.
Moreover, the unsuccessful conditions (in the $m^2\neq 0$ case) to
interrupt the chains, are
\begin{eqnarray}
\{G^{(0)}(x),H_0\}=m^2(\Pi^{nn}(x)-2\partial_nk_{0n}(x)) \,\, ,
\label{eqb40}
\end{eqnarray}
\begin{eqnarray}
\{G^{(0)}_i(x),H_0\}=m^2(\partial_nk_{in}(x)+\partial_ik_{00}(x)-\partial_ik_{nn}(x))
\,\, , \label{eqb41}
\end{eqnarray}
where we have fixed
\begin{eqnarray}
a(x,y)=0 \,\, , \label{eqb42}
\end{eqnarray}
\begin{eqnarray}
b^i(x,y)=-2\partial^i\delta^2(x-y) \,\, , \label{eqb43}
\end{eqnarray}
\begin{eqnarray}
a^i(x,y)=0 \,\, , \label{eqb44}
\end{eqnarray}
\begin{eqnarray}
{b^i}_j(x,y)=0 \,\, . \label{eqb45}
\end{eqnarray}

All this indicates that in the case where $m^2\neq 0$ there is not
a first class consistent chain of generators and, then there is no
gauge symmetry.

However, if we revisit the case $m^2=0$, conditions (\ref{eqb40})
and (\ref{eqb41}) are zero and the chains are interrupted. Now,
the generators $G^{(1)}$, $G^{(1)}_i$, $G^{(0)}$ and $G^{(0)}_i$
are first class. Using (\ref{eqb42})-(\ref{eqb43}), the generators
are rewritten again
\begin{equation}
G^{(1)}\equiv\Pi\,\, , \label{eqb46}
\end{equation}
\begin{equation}
G^{(1)}_i\equiv \partial_n
k_{ni}-\epsilon_{in}\partial_nv_0-\frac{\epsilon_{in}}{4}P^n+\frac{1}{4}\Pi^i\,\,
, \label{eqb47}
\end{equation}
\begin{eqnarray}
G^{(0)}=-\frac{\epsilon_{nm}}{2}\partial_nP^m-\frac{\partial_n\Pi^n}{2}
\,\, , \label{eqb48}
\end{eqnarray}
\begin{eqnarray}
G^{(0)}_i=-\frac{\partial_n\Pi^{ni}}{2}+\frac{\epsilon_{in}}{4}\partial_nP-\frac{\epsilon_{in}}{2}\Delta
v_n
-\frac{\epsilon_{in}}{2}\partial_n\partial_mv_m-\frac{\epsilon_{nm}}{2}\partial_i\partial_nv_m+\Delta
k_{0i}\,\, . \label{eqb49}
\end{eqnarray}

Introducing the parameters $\varepsilon (x)$ and $\varepsilon^i
(x)$, a combination of (\ref{eqb46})-(\ref{eqb49}) is taken into
account in the way that the gauge generator is
\begin{eqnarray}
G(\dot{\varepsilon} , \dot{\varepsilon}^i , \varepsilon ,
\varepsilon^i)=\big<
\dot{\varepsilon}(x)G^{(1)}(x)+\dot{\varepsilon}^i(x)G^{(1)}_i(x)+\varepsilon(x)
G^{(0)}(x) +\varepsilon^i(x)G^{(0)}_i(x)\big>\,\, , \label{eqb50}
\end{eqnarray}
and with this, for example the field transformation rules (this
means,  $\delta (...)= \{(...), G \}$) are written as
\begin{eqnarray}
\delta k_{00}= \dot{\varepsilon}\,\, , \label{eqb51}
\end{eqnarray}
\begin{eqnarray}
\delta k_{0i}=
\frac{\dot{\varepsilon}^i}{4}+\frac{\partial_i\varepsilon}{2}\,\,
, \label{eqb52}
\end{eqnarray}
\begin{eqnarray}
\delta k_{ij}=
\frac{1}{4}(\partial_i\varepsilon_j+\partial_j\varepsilon_i)\,\, ,
\label{eqb53}
\end{eqnarray}
\begin{eqnarray}
\delta v_0= \frac{\epsilon_{nm}}{4}\partial_n\varepsilon_m\,\, ,
\label{eqb54}
\end{eqnarray}
\begin{eqnarray}
\delta v_i= \frac{\epsilon_{in}}{4}\dot{\varepsilon}_n
-\frac{\epsilon_{in}}{2}\partial_n\varepsilon\,\, , \label{eqb55}
\end{eqnarray}
and, redefining parameters as follows: $\varepsilon \equiv 2\xi_0$
and $\varepsilon^i= 4 \xi^i$, it is very easy to see that these
rules match with (\ref{eqb9}) and (\ref{eqb10}), as we expected.

\vskip .2truein
\section{Concluding remarks}

A perturbative regime based on arbitrary variations of the
contorsion and metric as a (classical) fixed background, is
performed in the context of a pure Yang-Mills formulation of the
$GL(3,R)$ gauge group. There, we analyze in detail the physical
content and the well known fact that a variational principle based
on the propagation of torsion (contorsion), as dynamical and
possible candidate for a quantum canonical description of gravity
in a pure YM formulation gets serious difficulties.

In the $2+1$ dimensional massless case we show that the theory
propagates three massless degrees of freedom, one of them a
non-unitary mode. Then, introducing appropiate quadratical terms
dependent on torsion, which preserve parity and general
covariance, we can see that the linearized limit do not reproduces
an equivalent pure Hilbert-Einstein-Fierz-Pauli massive theory for
a spin-2 mode and, moreover there is other non-unitary modes.
Roughly speaking, at first sight one can blame it on the kinetic
part of YM formulation because the existence of non-positive
Hamiltonian connected with non-unitarity problem. Nevertheless
there are other possible models of Gauss-Bonnet type which could
solve the unitarity problem.

As we have given a glimpse, without introduction of explicit
massive $T^2$-terms in the action, it is possible to breakdown the
residual gauge invariance (i.e.: reduction of general covariance
which survives after a perturbative procedure) as a consequence
rising from the choice of a particular fixed non-Riemannian
background. The question of the existence of the alleged
geometrical mechanism which gives mass to the fields and the
existence of extensions of YM model which cure non-unitarity(i.e.: a possible family
of Gauss-Bonnet models with
torsion), will be explored elsewhere.

\vspace{5pt}

\noindent{\bf Aknowlegments}

Author thanks A. Restuccia for observations.


\end{document}